\documentclass[runningheads]{llncs}
\usepackage[T1]{fontenc}
\usepackage{graphicx,verbatim}
\usepackage{amsmath} 
\usepackage{booktabs}
\usepackage{tabularx}
\usepackage{multirow}
\usepackage{amssymb}

\usepackage{subfigure}
\begin{document}
\title{Cascaded 3D Diffusion Models for Whole-body 3D 18-F FDG PET/CT synthesis from Demographics}
%
\begin{comment}  %% Removed for anonymized MICCAI 2025 submission
\author{First Author\inst{1}\orcidID{0000-1111-2222-3333} \and
Second Author\inst{2,3}\orcidID{1111-2222-3333-4444} \and
Third Author\inst{3}\orcidID{2222--3333-4444-5555}}
%
\authorrunning{F. Author et al.}
\institute{Princeton University, Princeton NJ 08544, USA \and
Springer Heidelberg, Tiergartenstr. 17, 69121 Heidelberg, Germany
\email{lncs@springer.com}\\
\url{http://www.springer.com/gp/computer-science/lncs} \and
ABC Institute, Rupert-Karls-University Heidelberg, Heidelberg, Germany\\
\email{\{abc,lncs\}@uni-heidelberg.de}}

\end{comment}

\author{Siyeop Yoon\textsuperscript{*,a} \and Sifan Song\textsuperscript{*} \and Pengfei Jin \and Matthew Tivnan \and Yujin Oh \and Sekeun Kim \and Dufan Wu \and Xiang Li \and Quanzheng Li\textsuperscript{+,b}
}  %% Added for anonymized MICCAI 2025 submission
\authorrunning{Siyeop Yoon  et al.}
\institute{Massachusetts General Hospital and Harvard Medical School \\
    $^\textbf{*}$Co-first authors,$^\textbf{+}$Corresponding author\\  $^\text{a}$\email{syoon5@mgh.harvard.edu}, 
    $^\text{b}$\email{li.quanzheng@mgh.harvard.edu} 
}

\maketitle              % typeset the header of the contribution
\begin{abstract}
We propose a cascaded 3D diffusion model framework to synthesize high-fidelity 3D PET/CT volume directly from demographic variables, addressing the growing need for realistic digital twins in oncologic imaging, virtual trials, and AI-driven data augmentation. Unlike deterministic phantoms, which rely on predefined anatomical and metabolic templates, our method employs a two-stage generative process: an initial score-based diffusion model synthesizes low-resolution PET/CT volumes from the demographic variables only, providing global anatomical structures and approximate metabolic activity, followed by a super-resolution residual diffusion model refining spatial resolution. Our framework was trained on 18-F FDG PET/CT scans from the AutoPET dataset and evaluated using organ-wise volume and standardized uptake value (SUV) distributions, comparing synthetic and real data between demographic subgroups. The organ-wise comparison demonstrated strong concordance between synthetic and real images. In particular, most of the deviations in metabolic uptake values remained within 3–5\% of the ground truth in sub-group analysis. These findings highlight the potential of cascaded 3D diffusion models to generate anatomically and metabolically accurate PET/CT images, offering a robust alternative to traditional phantoms and enabling scalable, population-informed synthetic imaging for clinical and research applications.
% Authors must provide keywords and are not allowed to remove this Keyword section.

\end{abstract}

\section{Introduction}
Medical image synthesis has become an essential tool in healthcare, facilitating data augmentation \cite{akrout2023diffusion}, modality translation \cite{kim2024adaptive,wang2023patch}, and digital twin simulations \cite{kadry2024probing}. Generative models enable the creation of realistic medical images, addressing challenges related to data scarcity and privacy concerns \cite{kazeminia2020gans,pianykh2020continuous}. While early deep generative models such as VAEs \cite{kingma2013auto} and GANs \cite{goodfellow2020generative} demonstrated promise in medical image synthesis \cite{hu2022domain,pinaya2023generative}, diffusion models have recently emerged as the superior alternative, offering enhanced image fidelity and greater control over the generation process \cite{song2020score,ho2020denoising,EDM}. However, their high computational cost and slow inference speed remain significant challenges. To mitigate these computational demands, Latent diffusion models (LDMs) alleviate memory constraints by performing synthesis in a compressed latent space \cite{rombach2022high}, though the final image quality is heavily dependent on the encoder-decoder network used for latent mapping \cite{khader2023denoising}. In 3D volumetric image synthesis, patch-wise training has gained popularity \cite{wang2023patch,yoon2024high}. By breaking a volume into smaller patches, models can operate more efficiently while maintaining high-resolution outputs\cite{yoon2024volumetric}. Most previous studies on diffusion model-based image synthesis have focused on reconstructing images from undersampled measurements \cite{yoon2024high,chung2022score,jiang2023pet} or performing modality conversion using acquired images \cite{wang2024take,pan2024synthetic}. However, when models are weakly conditioned or operate unconditionally—without complementary imaging data, the iterative noise removal process often fails to maintain anatomical consistency, resulting in spatial incoherence \cite{pinaya2022brain}. 

The spatial contexts and alignment become even more challenging when models trained on different imaging modalities—such as anatomical and functional images—operate separately. Text-based image generation techniques have been explored as an alternative, using natural language descriptions to synthesize diverse medical images \cite{hamamci2024generatect,zhu2023make}. While such models can produce visually plausible results, they often lack anatomical precision due to the absence of explicit spatial constraints, leading to outputs that may not accurately reflect human anatomy.

To overcome these limitations, we propose an alternative framework that integrates stepwise 3D PET/CT volume synthesis with resolution enhancement, thereby overcoming challenges inherent in weakly conditional synthesis methods. Our approach first generates a coarse anatomical representation using only demographic attributes, establishing spatial relationships and organ layouts in a low-resolution 3D PET/CT volume. This blueprint is then progressively refined using a separated super-resolution residual diffusion model. By decoupling structural generation from visual enhancement, our method reduces dependence on large imaging datasets while streamlining the synthesis pipeline. The proposed framework is evaluated through task-based metrics, including organ volume accuracy and standardized uptake value (SUV) distributions from the AutoPET dataset. Our results confirm that the synthetic images closely align with real demographic-matched data, demonstrating high anatomical fidelity. 

\section{Methods}
\begin{figure*}[t!]
    \centering
    \includegraphics[width=\linewidth]{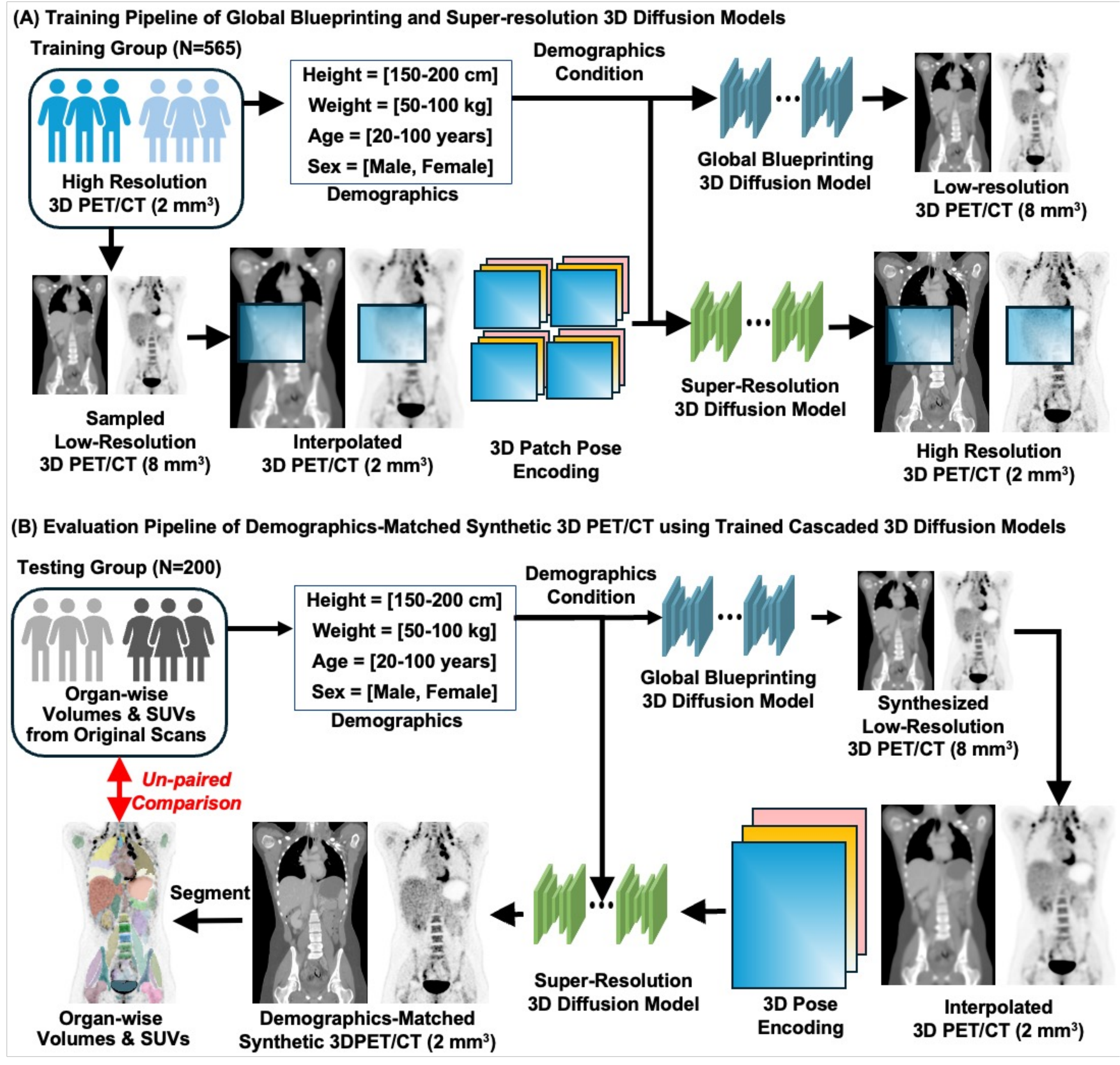}
    \caption{Overview of the cascaded 3D diffusion framework for demographic-driven PET/CT synthesis. 
\textbf{(A)}~During training, demographics guide the global diffusion model to generate a low-resolution 3D PET/CT, which is then interpolated and refined by a super-resolution diffusion model to produce high-resolution outputs. 
\textbf{(B)}~For evaluation, the demographics from the testing set are matched and input to the same cascaded process. This yields synthetic PET/CT compared to real data cohorts, focusing on organ-wise volume and SUV metrics.}
    \label{fig:enter-label}
\end{figure*}

We propose a cascaded 3D diffusion model framework that employs a global-to-local synthesis strategy to generate anatomically and metabolically consistent PET/CT images only from demographics (Fig. \ref{fig:enter-label}). 

\subsection{Cascaded Synthesis via Conditional Diffusion}

\subsubsection{Global Anatomical and Functional Synthesis}

In the first stage of our cascaded framework, a globally coherent low-resolution PET/CT volume is generated through a score-based diffusion process conditioned on demographic attributes. To formalize this, we define a family of distributions \(p({I_{\text{LR}}}; \sigma, x_{\text{Demo}})\) obtained by perturbing a low-resolution image with i.i.d. Gaussian noise of standard deviation \(\sigma\) with demographics $x_{\text{Demo}}$. At the maximum noise level \(\sigma_{\max}\), the distribution $p({I_{\text{LR}}}; \sigma_{\max}, x_{\text{cond}})$ approximates pure Gaussian noise, which serves as the initialization for the denoising procedure. The evolution from a noisy sample \(I_{\text{LR}}\) to a clean, structured, low-resolution volume is modeled by a stochastic differential equation (SDE) that combines deterministic drift with stochastic diffusion by \cite{anderson1982reverse}:
\begin{equation}
    \label{eq:sde_lowres}
    dI_{\text{LR}} = -\frac{1}{2}\beta(t) \, \nabla_{I_{\text{LR}}} \log p(I_{\text{LR}}; \sigma(t), x_{\text{Demo}})\, dt + \sqrt{\beta(t)} \, dW_t,
\end{equation}
where \(\beta(t)\) is a drift coefficient and \(dW_t\) represents a standard Wiener process. And \(\nabla_{I_{\text{LR}}} \log p(I_{\text{LR}}; \sigma(t), x_{\text{Demo}})\) denotes the score function that guides \(I_{\text{LR}}\) toward regions of higher probability density as the noise diminishes under the condition \(x_{\text{Demo}}\).  Numerical integration of the corresponding reverse-time ODE from Eq.~\eqref{eq:sde_lowres}—yields a stable and deterministic trajectory from a highly perturbed state to the final low-resolution PET/CT image. 

To approximate the score function \(\nabla_{I_{\text{LR}}} \log p(I_{\text{LR}}; \sigma(t), x_{\text{Demo}})\), we train a conditional score network \(s_\theta(\cdot, \sigma, x_{\text{Demo}})\) using the following loss:
\begin{equation}
    \label{eq:score_loss_lowres}
    \mathcal{L}_{\text{score, LR}} = \mathbb{E}_{\sigma, I_{\text{LR}}} \Bigl[ \lambda(\sigma) \, \bigl\| s_\theta(I_{\text{LR}} + \sigma \epsilon, \sigma, x_{\text{Demo}}) - I_{\text{LR}} \bigr\|^2 \Bigr],
\end{equation}
where \(\epsilon \sim \mathcal{N}(0,I)\) and \(\lambda(\sigma)\) is a weighting function that balances the contribution of different noise levels. Multiplication $\sigma$ to $\epsilon$ is to normalize the magnitude of the noise. This scaling allows the loss to remain stable across different noise levels and ensures that the network can generalize its denoising capability across the entire noise schedule. The generation of low resolution volume $I_{\text{LR}}$ is then evolved by integrating the reverse-time ODE:
\begin{equation}
  \label{eq:reverse_ode_lowres_sampling}
  \frac{dI_{\text{LR}}}{dt} = -\frac{1}{2}\beta(t) \, \nabla_{I_{\text{LR}}} \log p(I_{\text{LR}}; \sigma(t), x_{\text{Demo}}),
\end{equation}
from \(t = T\) (high noise) down to \(t = 0\) (no noise). This integration can be performed using standard numerical solvers (EDM2 Solver \cite{EDM,karras2024analyzing}) that provide a stable and accurate approximation of the trajectory. With the trained score network \(s_\theta(\cdot, \sigma, x_{\text{Demo}})\), the iterative denoising process reconstructs the complete anatomical and functional features, establishing a structural blueprint for the subsequent high-resolution synthesis stage.

\subsubsection{Super-Resolution via Residual Diffusion and Patch-Wise Training}
In the second stage, the low-resolution image \(I_{\text{LR}}\) obtained from the global synthesis is refined to recover fine anatomical details.  First, an upsampled estimate \(I_{\text{LU}}\) is computed by linear interpolation to the target high-resolution dimensions. Since simple interpolation does not fully restore the textures and edges, we define a residual term, $R = I_{\text{HR}} - I_{\text{LU}}$, where \(I_{\text{HR}}\) is the true high-resolution image.

Given that the low-resolution volume serves as a structural blueprint, patch-wise training can be employed effectively. The model is conditioned on the noise level as well as on a spatial prior that specifies the 3D location of each patch within the interpolated PET/CT volume,\(I_{\text{LU}}\). The 3D domain \(\Omega\) is partitioned into \(N\) patches, \(\Omega = \bigcup_{i=1}^N \Omega_i\), and the patch-wise loss is formulated as 
\begin{equation}
    L_{Patch} = \sum_{i=1}^N \mathbb{E}_{R, I_{\text{LU}},\sigma} \Bigl\| S_\theta\Bigl(R_{\Omega_i}+ \sigma \epsilon; I_{\text{LU},\Omega_i}),\sigma, x_{\text{Demo}}, \Omega_i\Bigr) - R_{\Omega_i} \Bigr\|^2_2
\end{equation}
, where \(R_{\Omega_i}\) denotes the residual signal corresponding to the positional encoding of patch \(\Omega_i\). \(I_{\text{LU}}\) and \(x_{\text{Demo}}\) represent the low-resolution PET/CT and demographics, respectively. The inclusion of the spatial \(\Omega_i\) and the low-resolution priors \(I_{\text{(LU, \(\Omega_i\)})}\) ensures that local reconstructions are consistent with the blueprint. Once the conditional score network \(S_\theta\) is trained with the above patch-wise loss, it is used to estimate the residuals for high-resolution PET/CT volumes through the reverse-time ODE flow, as defined in Eq.~\eqref{eq:reverse_ode_lowres_sampling}. Finally, the super-resolved high-resolution PET/CT volume is obtained by adding the recovered residual \(R\) to the upsampled estimate $ I_{\text{SR}} = I_{\text{LU}} + R$.
\begin{figure}[t]
    \centering
    \includegraphics[width=\linewidth]{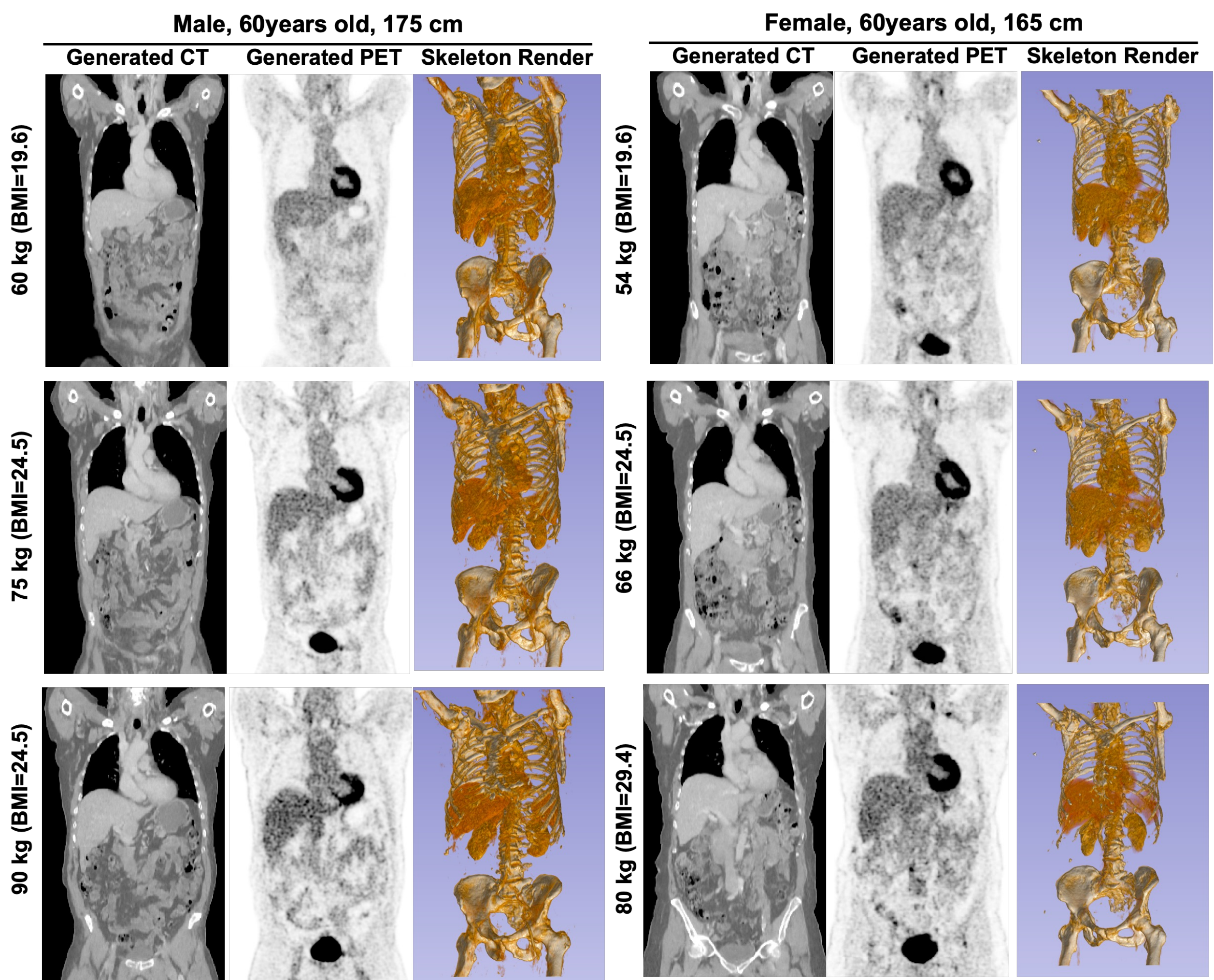}
    \caption{Representative examples of 18-F FDG PET/CT generated using cascaded 3D diffusion models. The images show CT, 18-FDG SUV, and 3D renderings for synthetic subjects of the same age (60 years) with different heights (male 175\,cm vs. female 165\,cm) and BMIs. The results demonstrate plausible anatomical and metabolic differences, including variations in adipose tissue distribution and PET signal heterogeneity.}
    \label{fig:BMIs}
\end{figure}
\subsection{Dataset and Implementation Details}
We utilize the AutoPET dataset, which comprises whole-body \textsuperscript{18}F-FDG PET/CT scans. A total of 765 subjects with complete demographic information (age, sex, height, and weight) were retrospectively selected (565 training and 200 testing subjects). All PET and CT were resampled to a voxel spacing of 2 mm\(^3\). 

To ensure consistent anatomical coverage, including key organs such as the heart, liver, and kidneys, CT volumes were segmented using the TotalSegmentator, and images were cropped to the region extending from the clavicle to the sacrum bones. Each volume was zero-padded to maintain a standardized spatial dimension: $ I_{\text{HR}} \in \mathbb{R}^{224 \times 224 \times 384}$. The low-resolution volumes were derived from the high-resolution volumes using voxel subsampling. Specifically, each HR volume was sampled by a factor of 4 along each spatial dimension using stratified voxel extraction, defined as $I_{\text{LR}}^{(x,y,z)} = I_{\text{HR}}[x::4, y::4, z::4]$, with Cartesian multiplication of $(x,y,z) \in \{0,1,2,3\}$. This process generates 64 unique low-resolution volumes per subject, significantly augmenting the training set size to low-resolution volumes while preserving anatomical diversity. 

To standardize intensity values, CTs were clipped in the range \([-500, 500]\) HU and linearly scaled to \([0, 1]\). PET images were converted to SUV units, clipped to \([0, 25]\), and normalized using $SUV_{\text{log}} = \frac{\log(SUV+1)}{\log(26)}$. During evaluation, normalized images were inverted to the original scale. Demographics were encoded as continuous values (age, height, weight), with sex encoded as a binary value.

For the model implementation, we extend the EDM2 framework \cite{karras2024analyzing,EDM} to the 3D domain for conditional PET/CT synthesis by integrating 3D convolutions. Our modified 3D U-Net captures spatial dependencies across all axes, ensuring anatomical fidelity while incorporating demographic attributes to enhance realism. The raining was performed on 16,252K images with an accumulated batch size of 2048, 64 base channels, and an initial learning rate of 0.017 decayed over 35,000 batches, requiring 72 hours on an NVIDIA DGX A100 system (4$\times$40GB A100 GPUs for each model). A two-stage approach is employed: a global context model using a $56\times56\times96$ input and a super-resolution model refining patches of the same size cropped from high-resolution volumes. Our framework allows flexible selection of the generative model, including a 3D flow-matching model \cite{lipman2022flow,lipman2024flow}, alongside diffusion models with identical settings.

In the model testing, for both diffusion and 3D flow-matching models, the number of sampling steps was set to 35 for global synthesis and 100 for super-resolution. All models were evaluated on a single 40GB A100 GPU. The global synthesis stage required a peak of 2GB GPU memory and approximately 30 seconds per sample, whereas the super-resolution stage used up to 24GB GPU memory and took about 6 minutes per sample. Note that super-resolution was performed on partial volumes of size \(224 \times 224 \times 96\), which were subsequently concatenated along the z-axis to reconstruct the full high-resolution output. %The source code of the proposed framework will be publicly released. 

Performance evaluation of the proposed cascaded 3D diffusion framework was conducted through both task-oriented quantitative analyses. Organ volumes were measured using segmentation masks obtained from both original and generated CT images via TotalSegmentator \cite{wasserthal2023totalsegmentator}, while organ SUV\(_{\text{mean}}\) and SUV\(_{\text{max}}\) were computed using the corresponding liver, heart, and kidney masks. Notably, failures or poor-quality segmentations were not observed in this study.
\section{Results}
Figure \ref{fig:BMIs} shows representative CT, 18-F FDG PET, and the corresponding surface volume rendering of synthetic images. The synthetic PET/CT volumes exhibit visually realistic anatomical and metabolic features that are consistent with variations in demographic inputs. In particular, the generated images capture subtle morphological differences such as variations in adipose tissue distribution and heterogeneous PET signals across subjects.
\begin{figure}[t]
    \centering
    \begin{subfigure}
        \centering
        \includegraphics[width=\linewidth]{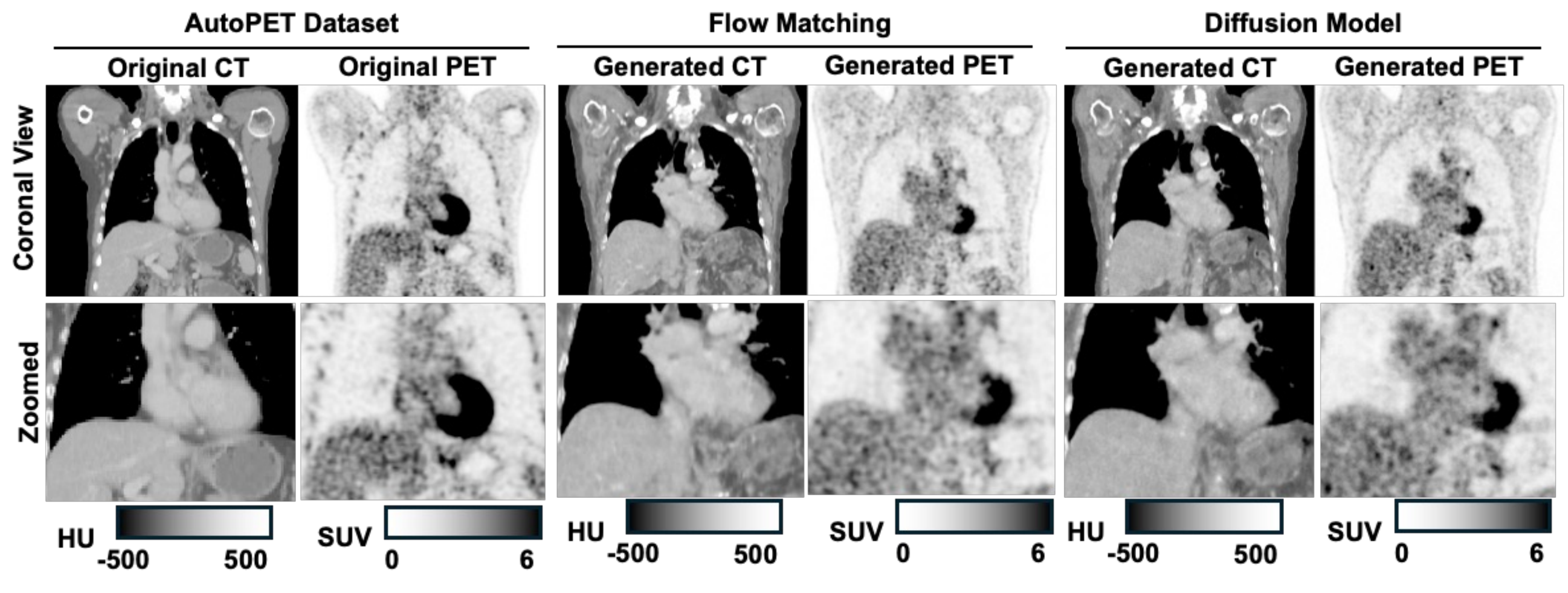}
        \label{fig:example1}
    \end{subfigure}
    \caption{Representative slices from the AutoPET (left) compared with synthetic CT/PET generated by the flow-matching (middle) and the diffusion model (right).}
\label{fig:combined}
\end{figure}

Table~\ref{tab:results} presents a quantitative comparison between the real (AutoPET) and synthetic datasets across multiple demographic groups. In general, synthetic PET / CT volumes showed strong concordance with real data at the group level. The mean SUV values in the synthetic PET images were within a close margin of the real data, and most subgroups did not show statistically significant differences (p>0.05) in liver and kidney volumes. However, female groups exhibited significant offsets in heart volume  (30 mL, 6.9\% difference). Similarly, a significant underestimation of heart SUV\(_{\text{max}}\) was observed in the male group. 

In our experiments, the flow-matching model produced synthetic PET/CT images with lower variability in quantitative metrics compared to the AutoPET data, as evidenced by a reduced standard deviation in organ volumes and SUV measurements. Flow matching using ODE formulations tends to yield less diverse samples compared to stochastic diffusion models, mainly because the deterministic integration in ODE approaches can constrain variability\cite{schusterbauer2024fmboost}. In contrast, diffusion models that rely on stochastic sampling can better capture and preserve the natural variability of the data, while the resolution enhancement stage yielded outputs with similar quality (Fig \ref{fig:example1}). Overall, shape metrics across demographic groups did not differ significantly, confirming that the synthetic images preserve anatomical structure while reflecting realistic metabolic activity.
\begin{table*}[t!]
\newcolumntype{C}{>{\centering\arraybackslash}X}
\centering
    \caption{Results of Organ-wise Volume and Standard Uptake Values (mean, peak) measured in the AutoPET and synthetic datasets of 18-FDG PET/CT. $^*$ indicates p-value <0.05 compared to AutoPET. Data are Mean ± Std (Mean difference \%).}
    \label{tab:results}
    \begin{tabularx}{\textwidth}{lccCCCC}\hline\hline
        \multicolumn{7}{c}{\textbf{Test cohort demographics}} \\ \hline
        & \multicolumn{3}{c}{\textbf{Male (N = 108)}} & \multicolumn{3}{c}{\textbf{Female (N = 92)}} \\ 
        \cmidrule(lr){2-4} \cmidrule(lr){5-7}
        
        %Participants &   & 108  &   &   & 92   &   \\
        Age (years) &   & 58 ± 17  &   &   & 59 ± 15   &   \\
        Heights (cm) &   & 177 ± 7&   &   & 165 ± 7  &   \\
        Weights (kg) &   & 83 ± 15 &   &   & 76 ± 19 &   \\\hline\hline
        \multicolumn{7}{c}{\textbf{Measurement in Liver}} \\ 
        %& \multicolumn{3}{c}{\textbf{Male} (N=108)} & \multicolumn{3}{c}{\textbf{Female} (N=92)} \\ 
        \cmidrule(lr){2-4} \cmidrule(lr){5-7}
        \textbf{Method} &  AutoPET & Flow & Diffusion & AutoPET & Flow & Diffusion \\ \hline

        \multirow{2}{*}{\shortstack{Volume(L)}} & 1.76±0.44 & 1.78±0.31 & 1.78±0.36 & 1.55±0.37 & 1.67±0.35$^*$  & 1.62±0.33 \\
                                           &     -    & (1.1\%)  & (1.3\%)  &    -     & (7.6\%) & (4.7\%)\\
        \multirow{2}{*}{\shortstack{SUV$_{\text{mean}}$}} & 2.26±0.42 & 1.96±0.26$^*$  & 2.18±0.34 & 2.34±0.33 & 2.15±0.22$^*$  & 2.38±0.40 \\
                                           &     -    & (-13.4\%) & (-3.9\%)  &    -     & (-8.6\%) & (1.5\%)\\
        \multirow{2}{*}{\shortstack{SUV$_{\text{max}}$}} &7.29±4.28& 5.27±1.83$^*$  &6.58±3.02  & 7.03±4.57    & 5.41±1.67$^*$  & 7.22±3.76    \\
                                           &     -    & (-27.7\%)  & (-9.7\%)  &    -     & (-23.1\%) & (2.8\%)\\\hline\hline

        \multicolumn{7}{c}{\textbf{Measurement in Heart}} \\ 
         \cmidrule(lr){2-4} \cmidrule(lr){5-7}
        %\textbf{Method} &  AutoPET & Flow & Diffusion & AutoPET & Flow & Diffusion \\ \hline
         \multirow{2}{*}{\shortstack{Volume(L)}} & 0.68±0.15 & 0.74 ± 0.10$^*$  & 0.71±0.11 & 0.55 0.09 & 0.63±0.09$^*$  & 0.58±0.09$^*$  \\
                                           &     -    & (8.3\%)  & (3.8\%)  &     -    & (15.8\%) & (6.9\%)\\
        \multirow{2}{*}{\shortstack{SUV$_{\text{mean}}$}} & 2.48±1.03 & 2.68±0.52 & 2.85±1.17 & 3.07±1.39 & 3.07±0.51 & 3.04±1.17 \\
                                           &     -    & (7.8\%) & (14.5\%)  &     -    & (0.2\%) & (-0.8\%)\\
        \multirow{2}{*}{\shortstack{SUV$_{\text{max}}$}} & 10.20±5.92 & 13.72±4.71$^*$  & 13.96±7.25$^*$  & 12.20±7.46 & 15.26±4.59$^*$  & 13.69±7.51 \\
                                           &     -    & (34.5\%)  & (36.8\%)  &    -     & (25.\%) & (12.2\%)\\\hline\hline
        
        \multicolumn{7}{c}{\textbf{Measurement in Kidneys}} \\ 
         \cmidrule(lr){2-4} \cmidrule(lr){5-7}
        %\textbf{Method} &  AutoPET & Flow & Diffusion & AutoPET & Flow & Diffusion \\ \hline
         \multirow{2}{*}{\shortstack{Volume(L)}} & 0.30±0.08 & 0.33±0.05$^*$  & 0.35±0.07 & 0.30±0.06 & 0.30±0.05 & 0.30±0.06 \\
                                           &     -    & (-6.9\%)  & (-2.3\%)  &    -     & (3.0\%) & (3.0\%)\\
        \multirow{2}{*}{\shortstack{SUV$_{\text{mean}}$}} &2.50±0.41 & 2.20±0.29$^*$  &2.47±0.42 & 2.70±0.45  & 2.48±0.17$^*$ & 2.71±0.38 \\
                                           &     -    & (-12.0\%) & (-0.9\%)  &    -     & (-8.1\%) & (0.4\%)\\
        \multirow{2}{*}{\shortstack{SUV$_{\text{max}}$}} & 11.20±5.24 & 8.90±2.77$^*$  & 10.63±4.04 & 11.83±5.55 & 11.39±3.58 & 11.79±3.63 \\
                                           &    -     & (-19.7\%)  & (-5.0\%)  &     -    & (-3.8\%) & (-0.4\%)\\\hline\hline
        
        \hline
    \end{tabularx} % Ensure column definitions and data match properly
\end{table*}

\section{Conclusion}
In this work, we introduced a cascaded 3D diffusion framework for the synthesis of PET/CT images directly from demographic variables. Our approach leverages a two-stage process, where an initial conditional diffusion model generates a coarse, low-resolution anatomical framework, and a subsequent super-resolution diffusion model refines this output to recover fine metabolic and structural details. This division of the synthesis task allows our method to effectively capture global anatomical structures while also ensuring that local, high-frequency details are faithfully reproduced.
Quantitative evaluation on the AutoPET dataset demonstrates that the synthetic images closely replicate key clinical metrics, including organ volumes and standardized uptake values (SUV), across a range of demographic groups. The strong agreement observed in liver, kidney, and heart measurements indicates that the framework is capable of producing anatomically accurate and metabolically realistic images.
The results of our study highlight the potential of this framework to serve as a reliable and scalable alternative to traditional imaging phantoms. By reducing the dependency on large, annotated imaging datasets, our method provides a novel solution for data augmentation, digital twin simulations, and virtual clinical trials. Future efforts will focus on optimizing the synthesis process for subpopulations where minor discrepancies persist and on extending the applicability of the method to additional imaging modalities. 

\bibliographystyle{splncs04}
\bibliography{Paper}

\end{document}